\documentclass[10pt]{article}
\usepackage{graphicx}

\newcommand{\beq}{\begin{equation}}
\newcommand{\eeq}{\end{equation}}

\def\sv{\left<\sigma v\right>}
\def\({\left(}
\def\){\right)}

\begin{document}

\title{Stable matter of 4th generation: hidden in the Universe and close to detection?}
%STABLE MATTER OF 4TH GENERATION: HIDDEN IN THE UNIVERSE AND CLOSE TO DETECTION?

\author{ K. Belotsky\footnote{e-mail: k-belotsky@yandex.ru},
M. Khlopov\footnote{e-mail: Maxim.Khlopov@roma1.infn.it},
K. Shibaev\footnote{e-mail: shibaev01@yandex.ru} \and 
{\small \it Moscow Engineering Physics Institute, 115409 Moscow, Kashirskoe sh.\ 31, Russia}}

\maketitle

\abstract{Stable neutrino and U quark of 4th
generation are excluded neither by experimental data, nor by
astrophysical constraints. Moreover, excess of stable $\bar U$
quarks in the Universe can lead to an exciting composite
nuclear-interacting form of dark matter, which can even dominate
in large scale structure formation.}

\section{Stable pieces of 4th generation matter}

The problem of existence of new families of quarks and leptons is
among the most important in the modern high energy physics. If
these quarks and/or leptons are stable, they should be present
around us and the reason for their evanescent nature should be
found.
Here we concentrate on the recent development of the approach
\cite{N}(see also \cite{Okun}, \cite{Q,I}), which involves minimal
number of parameters, but illustrates all the problems, related
with stable matter, composed of heavy charged particles.

Precision data on Standard model parameters admit \cite{Okun} the
existence of 4th generation
and the results of this analysis determine our choice for masses
of $N$ ($m_N=50$ GeV) and $U$ ($m_U=350 S_5$ GeV).

4th generation can follow from heterotic string phenomenology and
its difference from the three known light generations can be
explained by a new gauge charge ($y$-charge), possessed only by
its quarks and leptons \cite{N,Q,I}. Similar to electromagnetism
this charge is the source of a long range Coulomb-like
$y$-interaction. Strict conservation of $y$-charge makes the
lightest particle of 4th family (4th neutrino $N$) absolutely
stable, while the
lifetime of $U$ can exceed the age of the Universe, if $m_U<m_D$
\cite{Q,I}.

The following conditions fix $y$-charges ($Q_y$) of ($N,E,U,D$).
Cancellation of $Z-\gamma-y$ anomaly implies $Q_{yE} + 2\cdot
Q_{yU}+ Q_{yD} = 0$; while cancellation of $Z-y-y$ anomaly needs
$Q_{yN}^2 - Q_{yE}^2 +3\cdot(Q_{yU}^2 - Q_{yD}^2) = 0.$ Proper
$N-E$ and $U-D$ transitions of weak interaction assume $Q_{yN} =
Q_{yE}$ and $Q_{yU} = Q_{yD}$. From these conditions follows
$Q_{yE}=Q_{yN}=-3\cdot Q_{yU}=-3\cdot Q_{yD}$ so that $y$-charges
of ($N,E,U,D$) are ($1,1,-1/3,-1/3$).

$U$-quark can form lightest $(Uud)$ baryon and $(U \bar u)$ meson.
The corresponding antiparticles are formed by $\bar U$ with light
quarks and antiquarks. Owing to large chromo-Coulomb binding
energy ($\propto \alpha_{c}^2 \cdot m_U$, where $\alpha_{c}$ is
the QCD constant) stable double and triple $U$ bound states
$(UUq)$, $(UUU)$ and their antiparticles $(\bar U \bar U \bar u)$,
$(\bar U \bar U \bar U)$ can exist \cite{Q,I}. Formation of these
double and triple states in particle interactions at accelerators
and in cosmic rays is strongly suppressed, but they can form in
early Universe and strongly influence cosmological evolution of
4th generation hadrons. As shown in \cite{I}, \underline{an}ti-
\underline{U}-\underline{t}riple state called \underline{anut}ium
or $\Delta^{--}_{3 \bar U}$ is of special interest.

\section{Cosmological evolution of 4th generation matter}
In charge symmetric case, which assumes no primordial excessive
particles (or antiparticles) of 4th generation, cosmological
evolution was studied separately for $N$ and $U$ in \cite{N,Q,I}.

In the absence of particle (or antiparticle) excess primordial
particles of 4th generation with the considered masses can not
contribute significantly to total density. However, $N$ and $\bar
N$ condensation in galaxies can result in significant effect of
their annihilation in cosmic fluxes of positrons, antiprotons
 and in gamma background \cite{N}. They can even
provide simultaneous explanation of these data and results of the
direct WIMP searches \cite{N}, but in this case annihilation in
terrestrial matter of $N$ and $\bar N$, captured by Earth, should
lead to an up-going muon fluxes, exceeding the observed in
underground detectors.

The main problem of new stable quark is that it can form stable
positively charged hadrons.
In
charge symmetric case the pregalactic abundance of +2 charge
$(UUU)$ and $(UUu)$ is by 10 orders above the terrestrial limits
for anomalous helium ($r<10^{-19}$). For the abundance of +1
charge $(Uud)$ the situation is even more dramatic, since it
exceeds the limits on anomalous hydrogen ($r<10^{-30}$) by 20
orders of magnitude \cite{Q}.

To satisfy the experimental upper limits, anomalous isotope
abundance in Earth should be reduced. The mechanisms for such
reduction \cite{Q} imply recombination of $U$- and $\bar U$-
hadrons in dense objects of ordinary matter and their rapid
annihilation in bound $(U \bar U)$ states. These mechanisms
involve $y$-attraction, which prevents fractionating of $U$- and
$\bar U$- hadrons in matter. In the result the primordial
abundance of anomalous isotopes in terrestrial matter can be
reduced down to $r<10^{-28}$, so that the problem of anomalous
hydrogen can not be resolved in this way. Moreover, these
mechanisms, being sufficiently effective in the case of anomalous
helium, are accompanied by effects of energy release, which are
strongly constrained by the observed $\gamma$ background and by
the data from large volume detectors. All the above listed
problems of charge symmetric $N$ and $U$ matter can be avoided in
the case of primordial excess of $\bar U$ and $\bar N$.

The model  \cite{Q} admits that in the early Universe an
antibaryon asymmetry for 4th generation quarks can be generated
\cite{I}. Due to $y$-charge conservation $\bar U$ excess should be
compensated by $\bar N$ excess. $\bar U$-antibaryon density can be
expressed through the modern dark matter density $\Omega_{\bar U}=
k \cdot \Omega_{CDM}=0.224$ ($k \le 1$), saturating it at $k=1$.

In the early Universe at temperatures highly above their masses
$\bar U$ and $\bar N$ were in thermodynamical equilibrium with
relativistic plasma. It means that at $T>m_U$ ($T>m_N$) the
excessive $\bar U$ ($\bar N$) were accompanied by $U \bar U$ ($N
\bar N$) pairs.

Due to $\bar U$ excess frozen out concentration of deficit
$U$-quarks is suppressed at $T<m_U$ for $k>0.04$. It decreases
further exponentially,
when the frozen out $U$ quarks begin to bind with antiquarks $\bar
U $ into charmonium-like state $(\bar U U)$ and annihilate. On
this line $\bar U$ excess binds
by chromo-Coulomb
forces dominantly into $(\bar U \bar U \bar U)$ anutium states
with electric charge $Z_{\Delta}=-2$ and mass $m_o=1.05 S_5$TeV,
while remaining free $\bar U$ anti-quarks and anti-diquarks $(\bar
U \bar U)$ form after QCD phase transition normal size hadrons
$(\bar U u)$ and $(\bar U \bar U \bar u)$. Then at $T = T_{QCD}
\approx 150$MeV additional suppression of remaining $U$-quark
hadrons takes place in their hadronic collisions with $\bar
U$-hadrons, in which $(\bar U U)$ states are formed and $U$-quarks
successively annihilate.

Owing to weaker interaction, effect of $\bar N$ excess in the
suppression of deficit N is less pronounced but it still takes
place at $T<m_N$ for $k>0.02$. At $T \sim I_{NN} = \alpha_y^2
M_N/4 \sim 15 $MeV (for $\alpha_y = 1/30$ and $M_N=50$GeV) due to
$y$-interaction the frozen out $N$ begin to bind with $\bar N $
into charmonium-like states $(\bar N N)$ and annihilate. At $T <
I_{NU} = \alpha_y^2 M_N/2 \sim 30 $MeV $y$-interaction causes
binding of $N$ with $\bar U$-hadrons (dominantly with anutium) but
only at $T \sim I_{NU}/30 \sim 1$MeV this binding is not prevented
by back reaction of $y$-photo-destruction.

To the period of Standard Big Bang Nucleosynthesis (SBBN) $\bar U$
are dominantly bound in anutium $\Delta^{--}_{3 \bar U}$ with
small fraction ($\sim 10^{-6}$) of neutral $(\bar U u)$ and doubly
charged $(\bar U \bar U \bar u)$ hadron states. The dominant
fraction of anutium is bound by $y$-interaction with $\bar N$ in
$(\bar N \Delta^{--}_{3 \bar U})$ "atomic" state. Owing to early
decoupling of $y$-photons from relativistic plasma presence of
$y$-radiation background does not influence SBBN processes
\cite{Q,I}.

After $^4He$ is formed in SBBN, at $T \le T_{rHe} \sim
I_{o}/\log{\left(n_{\gamma}/n_{He}\right)} \approx I_{o}/27
\approx 60 $keV, where $I_{o} = Z_{\Delta}^2 Z_{He}^2 \alpha^2
m_{He}/2 \approx 1.6$MeV \cite{I}, neutral Anti-Neutrino-O-helium
(ANO-helium, $ANOHe$) $(^4He^{++} [\bar N \Delta^{--}_{3 \bar
U}])$ ``molecule'' with mass $m_o \approx 1S_5$TeV is produced in
the reaction $\Delta^{--}_{3 \bar U}+^4He\rightarrow \gamma
+(^4He^{++}\Delta^{--}_{3 \bar U}).$
The size of this
``molecule'' is $ R_{o} \sim 1/(Z_{\Delta} Z_{He}\alpha m_{He})
\approx 2 \cdot 10^{-13}$ cm
 and it can play the role of a dark matter component and
a nontrivial catalyzing role in nuclear transformations.

In nuclear processes ANO-helium looks like an $\alpha$ particle
with shielded electric charge. It can closely approach nuclei due
to the absence of a Coulomb barrier and opens the way to form
heavy nuclei in SBBN. This path of nuclear transformations
involves the fraction of baryons not exceeding $10^{-7}$ \cite{I}
and it can not be excluded by observations.

As soon as ANO-helium is formed, it catalyzes annihilation of
deficit $U$-hadrons and $N$. Charged $U$-hadrons penetrate neutral
ANO-helium, expel $^4He$, bind with anutium and annihilate falling
down the center of this bound system. The rate of this reaction is
$\sv= \pi R^2_o$ and an $\bar U$ excess $k=10^{-3}$ is sufficient
to reduce the primordial abundance of $(Uud)$ below the
experimental upper limits. $N$ capture rate is determined by the
size of $(\bar N \Delta)$ "atom" in ANO-helium and its
annihilation is less effective.

Interaction of the $^4He$ component of $(ANOHe)$ with a $^A_ZQ$
nucleus can lead to
formation of $^{A+4}_{Z+2}Q$ nucleus.
The final nucleus is formed in the excited $[\alpha, M(A,Z)]$
state, which can rapidly experience $\alpha$- decay, giving rise
to $(ANOHe)$ regeneration and to effective quasi-elastic process
of $(ANOHe)$-nucleus scattering. It leads to possible suppression
of ANO-helium catalysis of nuclear transformations in matter.

The composite nature of ANO-helium makes it more close to warm
dark matter. This dark matter plays dominant role in formation of
large scale structure at $k>1/2$.

The first evident consequence of the proposed scenario is the
inevitable presence of ANO-helium in terrestrial matter, which is
opaque for $(ANOHe)$ and stores all its in-falling flux. If its
interaction with matter is dominantly quasi-elastic, this flux
sinks down the center of Earth. If ANO-helium regeneration is not
effective and $\Delta$ remains bound with heavy nucleus $Z$,
anomalous isotope of $Z-2$ element appears. This is the serious
problem for the considered model.

Even at $k=1$ ANO-helium gives rise to less than 0.1 \cite{I} of
expected background events in XQC experiment \cite{XQC}, thus
avoiding for all $k \le 1$ severe constraints on Strongly
Interacting Massive particles SIMPs obtained in
\cite{McGuire:2001qj} from the results of this experiment. In
underground detectors $(ANOHe)$ ``molecules'' are slowed down to
thermal energies far below the threshold for direct dark matter
detection. However, $(ANOHe)$ destruction can result in observable
effects. Therefore a special strategy in search for this form of
dark matter is needed.

At $10^{-3}<k<0.02$ $U$-baryon abundance is strongly suppressed,
while the modest suppression of primordial $N$ abundance might not
exclude explanation of DAMA, HEAT and EGRET data in the framework
of hypothesis of 4th neutrinos \cite{N}, making the effect of $N$
annihilation in Earth consistent with the experimental data.

\section{Discussion}
Owing to excess of $\bar U$ anti-quarks primordial abundance of
$(Uud)$ is exponentially suppressed and anomalous isotope
over-production is avoided. Excessive anti-$U$-quarks should
retain dominantly in the form of anutium, which binds with
excessive $\bar N$ and then with $^4He$ in neutral ANO-helium.
Galactic cosmic rays destroy ANO-helium, striking off $^4He$. It
can lead to appearance of a free [anutium-$\bar N$] component in
cosmic rays, which can be as large as $[\bar N \Delta^{--}_{3 \bar
U}]/^4He \sim 10^{-7}$ and accessible to PAMELA and AMS
experiments. In the context of composite dark matter like \cite{I}
accelerator search for new stable quarks and leptons acquires the
meaning of critical test for existence of its charged components.
%
%\begin{figure}[ht]
%\begin{tabular}{p{6.5cm}p{4.4cm}}
%\special{isoscale crossec.bmp, 6.5cm 4.4cm} & \\ & {\footnotesize
%{\bf Figure 1.} Cross sections of production of 4th generation
%particles in experiment \mbox{ATLAS}\@. The level of X-axis
%corresponds to the lower (conservative) level of LHC sensitivity
%for 1st year of its operation.} \vspace{0.5cm}
%\end{tabular}
%\end{figure}
%
\begin{figure}[ht]
\begin{tabular}{rr}
%\centerline{\epsfxsize=4.5cm\epsfbox{crossec.eps}}
\ \ \ \includegraphics[width=5.5cm]{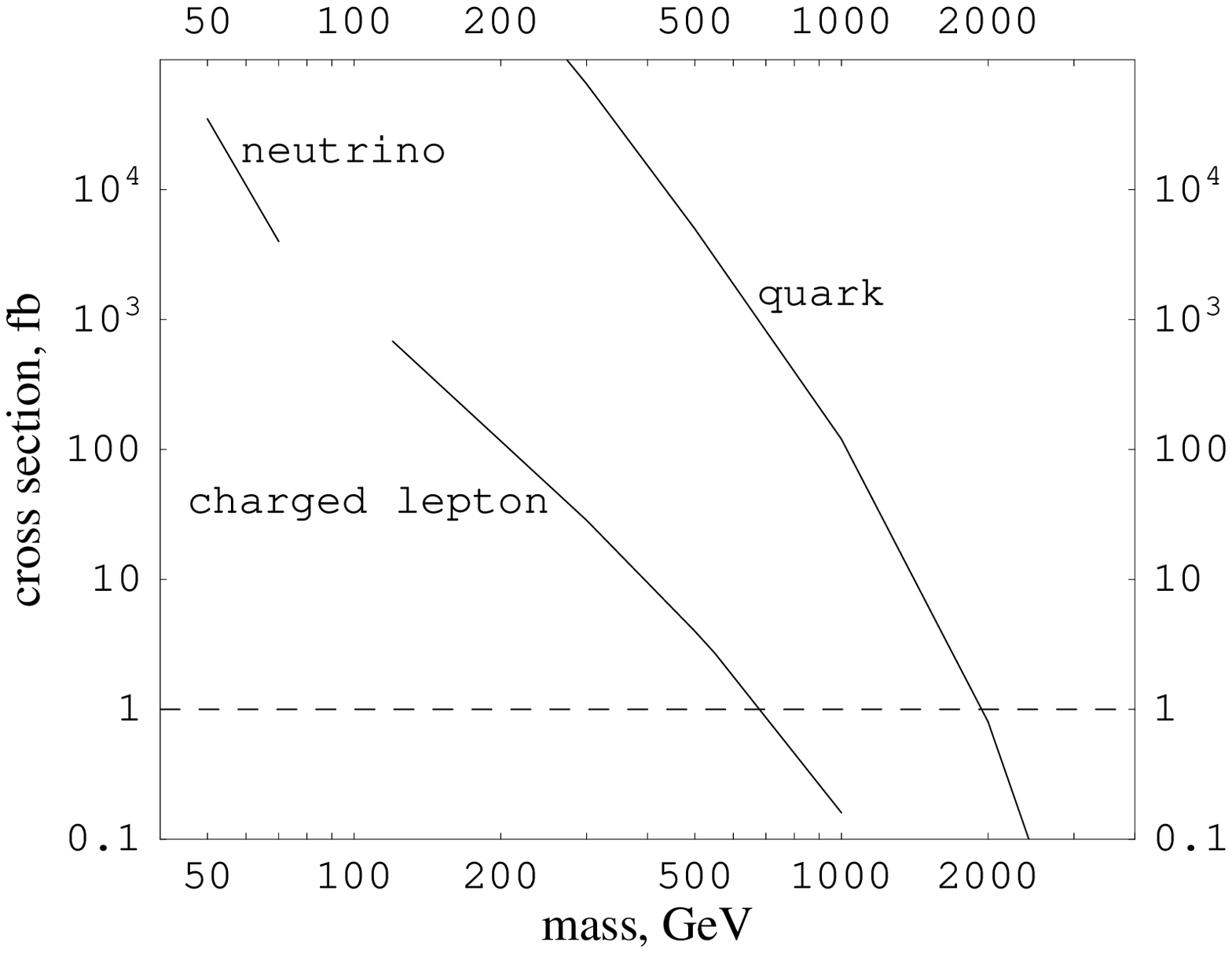} &
\raisebox{2.5cm}{\parbox{5cm}{\footnotesize {\bf Figure 1.} 
Cross sections of production of 4th generation particles 
in experiment \mbox{ATLAS}\@. Dash line shows the lower 
(conservative) level of LHC sensitivity for 1st year of its operation.}}
\end{tabular}
\end{figure}

\section*{Acknowledgments}

The work was supported by Khalatnikov-Starobinsky school.
M.Kh.\ thanks LPSC (Grenoble, France) for hospitality and D. Rouable for help.

\end{document}